\documentclass[]{emulateapj}

\slugcomment{}
\journalinfo{ApJL, accepted 2007/11/14}

\newcommand{\myemail}{pedro@ifa.hawaii.edu}
\newcommand{\tna} {\,\tablenotemark{a}}
\newcommand{\tnb} {\,\tablenotemark{b}}
\newcommand{\tnc} {\,\tablenotemark{c}}
\newcommand{\tnd} {\,\tablenotemark{d}}
\newcommand{\tne} {\,\tablenotemark{e}}

\shorttitle{Detection of Contact Binaries}
\shortauthors{Lacerda}

\begin{document}

\def\FigGeometry{
  \begin{figure}
    \centering
    \includegraphics[width=0.37\textwidth]{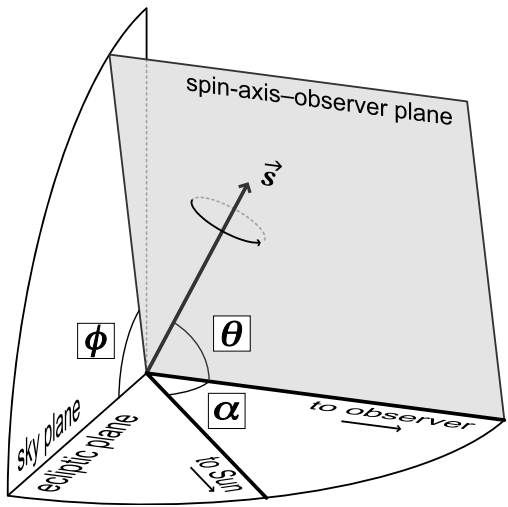}

    \caption[f1.eps] {Angles used to describe the observing geometry. The
orbital spin axis is denoted $\vec{s}$. The contact binary is assumed located
at the vertex where the angles $\theta$ (aspect, the angle between the orbital
spin axis and the line-of-sight), $\alpha$ (phase, the angular distance between
the Sun and the observer, as seen from the object), and $\phi$ (azimuth, the
sky-plane angle between the projections of the orbital spin axis and the
object-Sun vector) are measured.} 

	\label{Fig.Geometry}
  \end{figure}
}

\def\FigAlphaEff{
  \begin{figure}
    \centering
    \includegraphics[width=0.4\textwidth]{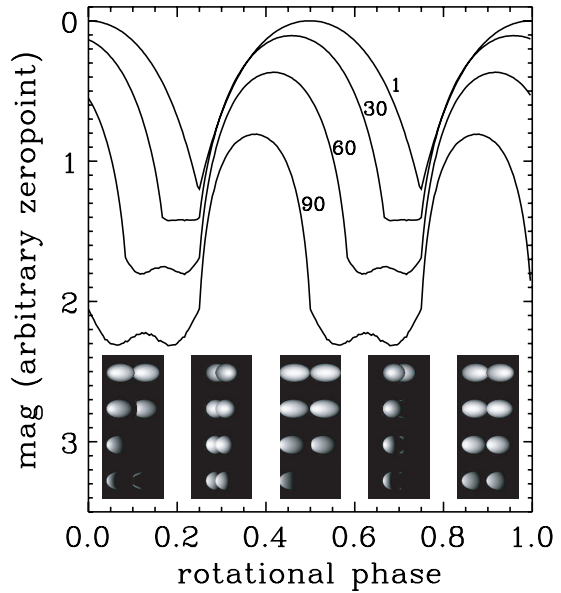}

    \caption[f2.eps] {Change in lightcurve morphology with increasing phase
angle. Lines are labelled with the corresponding phase angles ($\alpha=1\degr,
30\degr, 60\degr,$ and $90\degr$). Insets show rendering of the binary used to
produce the lightcurves (phase angle increasing from top to bottom) at
rotational phases $\phi=0.1, 0.3, 0.5, 0.7,$ and $0.9$.} 

	\label{Fig.AlphaEff}
  \end{figure}
}

\def\FigEffSparse{
  \begin{figure}
    \centering
    \includegraphics[width=0.45\textwidth]{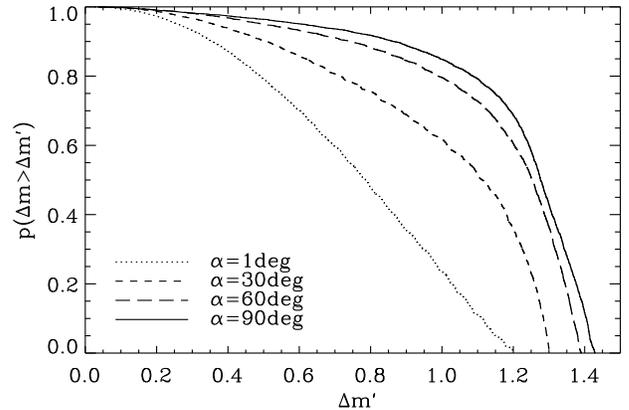}

    \caption[f3.eps] {Probability of measuring a magnitude range
larger than $\Delta m'$ from random sampling each lightcurve on
Fig.~\ref{Fig.AlphaEff} five times. An observing geometry $\theta=\phi=90\degr$
has been assumed (see text and Fig.~\ref{Fig.Geometry}).  } 

	\label{Fig.EffSparse}
  \end{figure}
}

\def\FigEffects{
  \begin{figure*}
    \centering
    \includegraphics[width=0.85\textwidth]{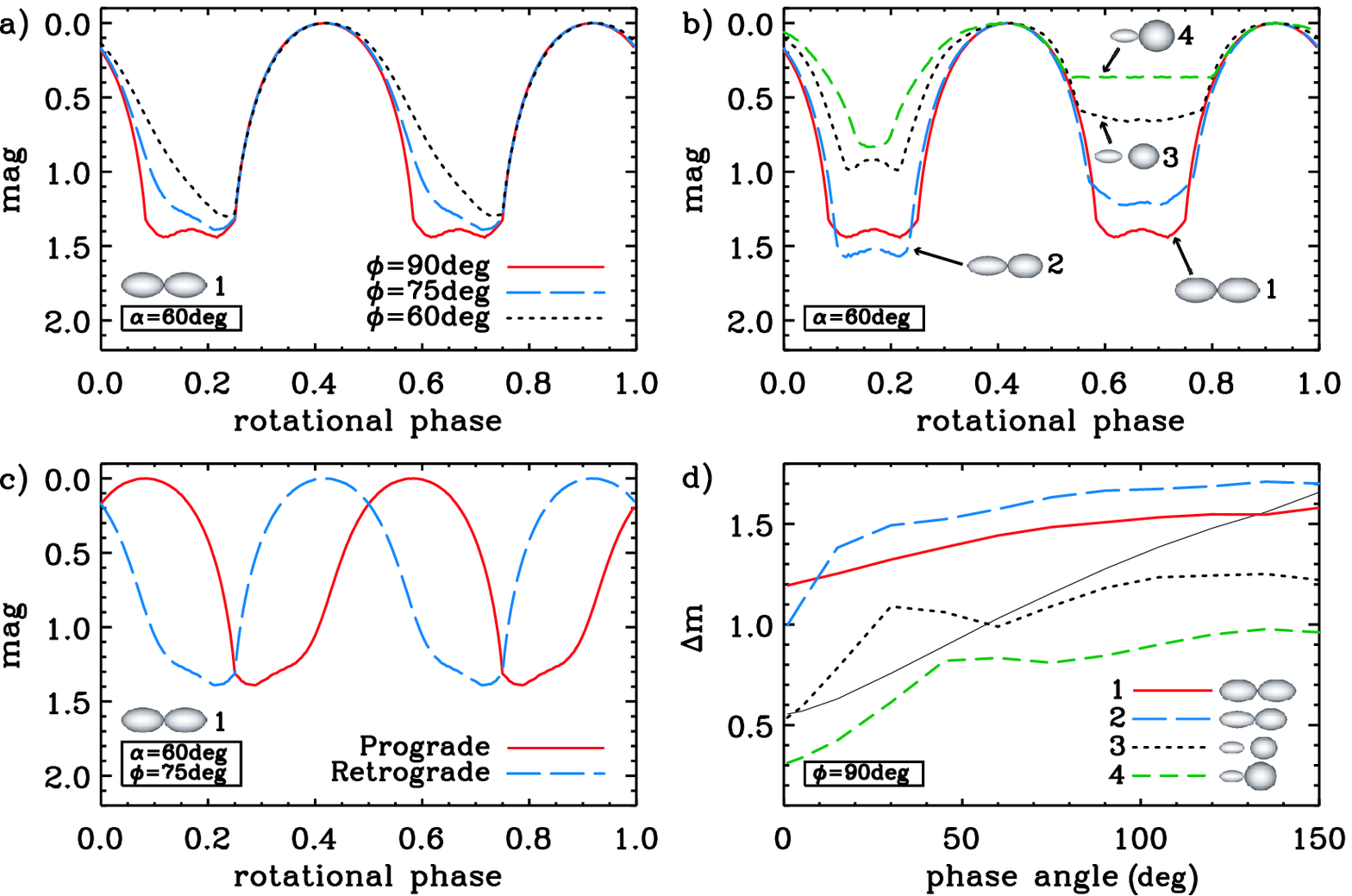}

    \caption[f4.eps] {{\it a)} Lightcurves of system \#1
(Table~1) as a function of azimuth angle $\phi$. Geometry is
$\alpha=60$\degr. {\it b)} Lightcurves of all four systems in
Table~1 at $\alpha=60$\degr\ and $\phi=90$\degr. {\it c)}
Asymmetric lighcurves of system \#1 rotating in the prograde and retrograde
directions. Asymmetry is due to observing geometry $\phi\neq90$\degr\ and high
phase angle $\alpha=60$\degr\ and shows clear dependence on the direction of
rotation of the binary. {\it d)} Lightcurve range versus phase angle
($\alpha=1,5,15,30,45,60,75,90,105,120,135,$ and 150\degr) for each system in
Table~1. Thin solid black line corresponds to a Jacobi
ellipsoid with axes ratios $b/a=0.60$ and $c/a=0.43$.  Observing geometry is
$\phi=90\degr$ (Fig.~\ref{Fig.Geometry}). All four plots assume
Lommel-Seeliger lunar-type surface scattering and aspect $\theta=90$\degr.} 

	\label{Fig.Effects}
  \end{figure*}
}

\title{Detection of Contact Binaries Using Sparse High Phase Angle Lightcurves}

\author{Pedro Lacerda}

%
\affil{Institute for Astronomy, University of Hawaii, 2680 Woodlawn
Drive, Honolulu, HI 96822}

\email{\myemail}

\begin{abstract}

We show that candidate contact binary asteroids can be efficiently identified
from sparsely sampled photometry taken at phase angles $\alpha>60$\degr. At
high phase angle, close/contact binary systems produce distinctive lightcurves
that spend most of the time at maximum or minimum (typically $>$$1\,$mag apart)
brightness with relatively fast transitions between the two.  This means that a
few ($\sim$5) sparse observations will suffice to measure the large range of
variation and identify candidate contact binary systems.  This finding can be
used in the context of all-sky surveys to constrain the fraction of contact
binary near-Earth objects. High phase angle lightcurve data can also reveal the
absolute sense of the spin.

\end{abstract}

\keywords{minor planets, asteroids --- solar system: general --- techniques:
photometric}

\section{Introduction}

For most purposes, small solar system bodies are best observed at solar
opposition. Measurements taken at low phase angle (the Sun-target-observer
angle, denoted $\alpha$; see Fig.~\ref{Fig.Geometry}) benefit from strong
backscattering, thus rendering the target brighter. In astrometric or
surface-averaged measurements, the enhanced signal-to-noise due to the low
phase angle geometry is desirable.  However, if the goal is to derive
information on the shape of the target, then high-$\alpha$ data become useful.
This has been noted in the past, in the context of lightcurve inversion
problems \citep{2003A&A...404..709D}. As a body rotates, the amount of sunlight
it reflects to the observer depends on the time varying projected
cross-section, which in turn depends on its shape.  If well-sampled lightcurve
information is obtained at multiple observing geometries, then the shape of the
object can be derived. At low phase angles the non-convex features of an object
are not apparent from its lightcurve.  This means that in the absence of
high-$\alpha$ data the non-convex figure of a contact binary can be wrongly
interpreted as being composed of a single convex component. 

One known effect of high-$\alpha$ observations on lightcurves is the increase
in the peak-to-peak range of variation \citep{1956ApJ...123..331G}.  For single
ellipsoidal bodies, and for phase angles $\alpha < 40\degr$, the lightcurve
range $\Delta m$ has been shown to vary roughly linearly with $\alpha$
\citep{1990A&A...231..548Z}. This is generally known as the amplitude-phase
relationship (APR). When observing from Earth, the maximum phase angle at which
a small solar system body can be imaged depends essentially on its heliocentric
distance.  The distant Kuiper belt objects can be observed up to
$\alpha\sim2\degr$, Jovian Trojan asteroids at $\alpha\lesssim10\degr$, and
main-belt asteroids at $\alpha\lesssim30\degr$.  Only near-Earth objects (NEOs)
can be observed at phase angles $\alpha>30\degr$ which makes them the best
candidates for high-$\alpha$ studies.

On the order of twenty binary NEOs have been discovered, both photometrically
\citep[e.g.,][]{1997Icar..127..431P,1998Icar..133...79P} and from radar
observations \citep[e.g.,][]{2000IAUC.7518....2N,2002aste.conf..151O}. From the
known sample, it is possible to note a few characteristic features
\citep{2006Icar..181...63P}.  The primaries have mean diameters of $\sim 1\,$km
(with a spread of one order of magnitude), usually two to four times those of
their secondary companions.  One case, 69230 Hermes, has a size ratio close to
unity \citep{2003IAUC.8227....2M}.  The separation distances are small, usually
a few primary radii, and the mutual orbits are nearly circular.  Most measured
primaries spin (around the shortest physical axis) close to the critical
break-up rate for strengthless bodies with mean density $\sim
2\,$g$\,$cm$^{-3}$. The secondaries have synchronized spin and orbital periods.
In the few ($\sim$10) cases where shapes can be inferred, the primaries are
well described by oblate spheroids, while the secondaries have elongated shapes
along the line of centers. One NEO, 4769 Castalia, is suspected from radar data
to be a contact binary \citep{1990Sci...248.1523O}.

In recent years, explanations to how binary NEOs may form have mostly converged
on the idea of the splitting of a parent body.  \citet{1980Icar...44..807W}
proposed that binaries may form by rotational fission of an object effected by
an off-center impact.  \citet{2002Sci...296.1445M} argued that spin-up due to
tidal interactions during close encounters with planets was the most likely
formation process for NEO binaries, and \citet{2006AREPS..34..157B} were the
first to suggest that radiaton forces (YORP effect,
Yarkovsky-O'Keefe-Radzievskii-Paddack) provide the torque needed for rotational
fission.  The main features of the NEO binary sample described above have led
\citet{2007ApJ...659L..57C} to propose the following formation mechanism: due
to YORP, $\sim 1\,$km objects are spun up close to break-up rates, which leads
to the accumulation of regolith around the equator. As YORP continues to spin
up the body, the surface material is eventually stored in orbit around the
(now) primary, where it coagulates into a secondary; the latter grows into an
elongated, nearly Roche equilibrium shape balancing gravitational and inertial
accelerations. The subsequent evolution of the mutual orbit is driven by BYORP,
the binary version of YORP, which may lead the components away from each other
but may also bring them together to mutual contact
\citep{2005Icar..176..418C,2007Icar..189..370S}.

To obtain a well-sampled lightcurve of an NEO, capable of unequivocally
establishing its nature, would require extensive observations at multiple
geometries. For a sparse survey such as will be undertaken by Pan-STARRS
(typically revisiting each target $\sim4$ times per lunation), it may take
years before the shape of the object can be determined. The method presented
here can single out potential contact binaries after $\sim5$ measurements at
high-$\alpha$.  In this letter, we present the first lightcurve simulations of
contact binaries at high phase angle, and use them to address the detectability
of such systems.

\FigGeometry

\begin{deluxetable}{ccccccc}
  \tablecaption{Four close/contact binary systems considered.}
    \label{Table.Binaries}
   \tablewidth{0pt}
   \tablehead{
   \colhead{\#\tna} & \colhead{$q$\tnb} & \colhead{($B/A, C/A$)\tnc} & \colhead{($b/a, c/a$)\tnc}  & \colhead{$l$\tnd} & \colhead{H.E.\tne}
   }
   \startdata
    1 & 1.00 & (0.66, 0.60) & (0.66, 0.60) & 1.00 & yes \\
    2 & 0.67 & (0.77, 0.69) & (0.53, 0.49) & 1.00 & yes \\
    3 & 0.25 & (0.92, 0.83) & (0.51, 0.48) & 1.19 & yes \\
    4 & 0.13 & (1.00, 0.90) & (0.51, 0.48) & 1.00 & no \\
   \enddata
  \tablenotetext{a}{System number}
  \tablenotetext{b}{Mass ratio}
  \tablenotetext{c}{Primary and secondary axis ratios}
  \tablenotetext{d}{Orbital distance in units of $A+a$}
  \tablenotetext{e}{Components in hydrostatic equilibrium}
\end{deluxetable}

\section{High Phase Angle Lightcurves}

The characteristic morphological features of contact binary lightcurves are
the large range of variation, typically 0.9$\,$mag or more, the rounded,
inverted-U shaped maxima, and the sharp V shaped minima
\citep{1980M&P....22..153Z,1984A&A...140..265L}. However, these traits appear
only at phase angles close to $\alpha=0\degr$. Figure \ref{Fig.AlphaEff} shows
lightcurves of a symmetric contact Roche binary (system \#1 from Table~1) at
four different phase angles, $\alpha=1$, 30, 60, and 90$\,\degr$. The surface
scattering of light is modelled using a Lommel-Seeliger function, taken to
represent a low albedo, lunar-type surface.  The procedure used for generating
the model lightcurves of Roche binaries, as well as of Jacobi triaxial
ellipsoids, is described in detail in \citet{2007AJ....133.1393L}.  
The most noticeable mutations as $\alpha$ increases include: 
\begin{enumerate}
\item the shape of the minimum changes from V-shaped first to flat and then to
slightly W-shaped; 
\item the transition between low and high brightness becomes sharper; 
\item the positions (in rotational phase) of the minima and maxima drift to the
left; 
\item the overall brightness decreases, as the illuminated fraction of the
surface diminishes; 
\item the trough-to-peak range of variation increases. 
\end{enumerate} 
Below, we discuss some of these points in more detail.

\subsection{Lightcurve Shape and Detection Probability}

As a consequence of points 1 and 2 above, the probability of detecting the
total range of brightness variation from only a few measurements increases
significantly with $\alpha$. This is shown in Fig.~\ref{Fig.EffSparse} where we
plot the probability of measuring the extent of variation of a contact binary
lightcurve for different values of $\alpha$.  The Figure was generated as
follows: each of the four lightcurves in Fig.~\ref{Fig.AlphaEff} (corresponding
to each $\alpha$) was sampled five times at random rotational phases, and the
maximum range $\Delta m_i$ within the set of five measurements was registered.
This procedure was then repeated for $i=1$ to 10000, and the cumulative
distribution of ranges $\Delta m_i$ was plotted for each $\alpha$, starting
from the maximum range. The plot shows that beyond $\alpha\sim 60\degr$ most
five-point samples of the lightcurve are able to identify a variation larger
than 1$\,$mag. For example, at $\alpha=60\degr$ there is $\sim$80\% chance of
detecting a $\Delta m\geq 1\,$mag from five sparse observations of the contact
binary considered.  

\FigAlphaEff

Figure~\ref{Fig.EffSparse} was obtained assuming the most favorable observing
geometry (Fig.~\ref{Fig.Geometry}): $\theta=90\degr$, measured between the spin
pole and the line-of-sight,  and $\phi=90\degr$, measured in the plane of the
sky between the spin pole and the object-Sun vector. The total range of a
contact binary lightcurve decreases with the aspect angle $\theta$ at about
0.03 to 0.04$\,$mag/$\degr$ \citep{2007AJ....133.1393L}; the maximum range is
obtained at $\theta=90\degr$.  The azimuthal angle $\phi$ also influences the
lightcurve range and shape. As $\phi$ moves away from 90$\degr$ (the direction
does not matter) the lightcurve becomes asymmetric and the range of variation
is slightly decreased (Fig.~\ref{Fig.Effects}a).  Therefore, requiring a
favorable geometry, i.e., angles larger than chosen minimum values
$\theta>\theta_\mathrm{m}$ and $\phi>\phi_\mathrm{m}$ reduces the probability
of detection of large $\Delta m$.  Assuming randomly oriented spin axes, the
probability of having simultaneously favorable $\theta$ and $\phi$ is given by
\begin{eqnarray}
 \lefteqn{1 - \frac{2}{4\,\pi }\,\Big[ \int _{0}^{\theta_\mathrm{m}} 
   \int_{0}^{2\,\pi }\sin (\theta )\,d\phi \,d\theta \, + } \\ \nonumber
& & \int_{\theta_\mathrm{m}}^{\frac{\pi }{2}}
  \int_{-\phi_\mathrm{m}}^{\phi_\mathrm{m}}\sin (\theta )\,d\phi \,d\theta \Big]
   = \frac{\left( \pi  - \phi_\mathrm{m} \right)}{\pi } \,\cos
(\theta_\mathrm{m}), 
\end{eqnarray}
which for minimum aspect and azimuth
$\theta_\mathrm{m}=\phi_\mathrm{m}=75\degr$ is $\sim0.15$. This probability
must be taken into account when estimating the intrinsic fraction of contact
binary systems from a sparsely sampled survey. At any rate, since the geometry
constraint (dominated by $\theta$) is present at all phase angles the
conclusion that high-$\alpha$ measurements are more effective at identifying
contact binary systems holds.

Figure \ref{Fig.Effects}b shows how the lightcurve shape depends on the
relative size and separation of the binary components. Four systems were
considered -- their properties are detailed in Table~1.  Except for system \#3,
meant to represent a system formed according to the model by \'Cuk mentioned in
\S1, all systems are Roche binaries in hydrostatic equillibrium. Systems \#3
and \#4 produce shallower lightcurves due to their asymmetric mass ratio, and
although the maxima and minima take up a large fraction of the rotational
phase, the probability of measuring the total range of variation is not as high
as for the more symmetric binaries \#1 and \#2. In conclusion, high-$\alpha$
observations will more easily detect contact binaries with similar sized
components.

Currently known binary NEOs are generally asymmetric with primary to secondary
size ratios of 2 to 4 \citep{2006Icar..181...63P}. Furthermore, no very close
pairs or contact binaries have been directly observed [with the possible
exception of 25143 Itokawa \citep{2007Icar..188..425S}]. The minimum orbital
period is $P=11\,$hr.  Whether these are intrinsic features or the result of
observational bias is still unclear, but the aptness of the method presented
here to find symmetric contact binaries should help throw light on the subject.
The intrinsic fraction of close and contact binaries is a powerful constraint
for models of NEO binary formation and destruction.

\FigEffSparse

\subsection{Lightcurve Phase Shift and Spin Direction}

Increasing the phase angle $\alpha$ produces a shift in the rotational phase of
the lightcurves (see Fig.~\ref{Fig.AlphaEff}). The shift happens respectively
to the left or right depending on the object being illuminated from the left or
right side, from the observer's standpoint. We measure a linear rotational
phase shift with a slope $\sim1.4\times10^{-3}$ per degree phase angle.  The
shift slope is similar for all four systems in Table~1. This effect must be
taken into account when fitting a single rotation period to data taken at
different phase angles if not to be confused with evidence for complex
rotation.

As mentioned in \S2.1, if the geometry is such that $\phi\neq90\degr$ then
asymmetries arise in the lightcurve. The direction of the asymmetry depends on
the sense of rotation of the binary. Figure~\ref{Fig.Effects}c illustrates this
effect. The asymmetry can thus be used to infer the sense of spin of the
binary.

\subsection{Lightcurve Range}

Careful inspection of Fig.~\ref{Fig.AlphaEff} shows that the lightcurve range
increases with $\alpha$. This is more easily seen in Fig.~\ref{Fig.Effects}d
where lightcurve range is plotted versus phase angle (APR) for each of the four
systems in Table~1. For comparison, a thin black solid line illustrates the
same dependence for a triaxial Jacobi ellipsoid with axis ratios $b/a=0.60$ and
$c/a=0.43$.

We find that, with the exception of the symmetric binary \#1, the APRs of
binary systems appear less regular than that of the ellipsoid. Binary APRs show
two roughly linear regimes, steeper for lower $\alpha$ and shallower for larger
$\alpha$. The slope only seems to depend on system type for small to
intermediate phase angles: beyond $\alpha\sim50$\degr\ the slopes are similar
for all systems. Systems with smaller ranges at $\alpha=1$\degr\ have shallower
initial APR slopes. This is similar to what was found by
\citet{1990A&A...231..548Z} for ellipsoids.  System \#1, however, has a
remarkably regular APR. The slope is 0.004$\,$mag/\degr\ up to
$\alpha\sim75$\degr\ and 0.001$\,$mag/\degr\ beyond that value. Incidentally,
the symmetric binary 90 Antiope \citep{2000DPS....32.1306M,2007Icar..187..482D}
has an APR slope 0.005$\,$mag/\degr\ \citep{2002A&A...396..293M}, very close to
that of our simulated symmetric binary.  The APR is shape-dominated: the APR
slopes do not depend significantly on the choice of 
surface scattering model. 

Our results seem to indicate that less regular (more asymmetric) objects have
steeper APRs. Our most regular object, the symmetric binary, has the shallowest
APR.  A detailed analysis of the APR of contact binary systems is beyond the
scope of this letter and will be treated elsewhere.

\FigEffects

\section{Summary}

We have presented the case for using high phase angle observations to find
extreme lightcurves produced by contact binaries in the solar system. Sparse
observations of contact binaries at phase angles $\alpha>60$\degr\ are
extremely efficient at detecting the large range of brightness variation
characteristic of these objects' lightcurves. From Earth, this method is most
relevant to observations of near-Earth objects as they attain the largest phase
angles.  At this point, a survey of high-$\alpha$ measurements alone should
probably not be used to estimate the fraction of contact binaries in the NEO
population.  Deviations from hydrostatic equilibrium shape and the precise
observing geometry ($\theta$ and $\phi$) affect the exact detection
probability, thereby introducing uncertainty in the derived abundance. The
technique is extremely efficient at detecting equator-on, relatively smooth
contact binaries with symmetric components, but confirmation of the exact
configuration of a system that shows large $\Delta m$ in just a few high-$\alpha$
measurements should always be sought using follow-up observations.  The sparse
quality of planned all-sky surveys such as Pan-STARRS implies that a full shape
solution can be obtained only after several years of data have been collected.
It is therefore extremely useful to develop fast techniques that allow the
identification of potentially interesting targets from a Pan-STARRS-type survey
(or even from dedicated surveys using sub-1m telescopes), which can be followed
up on other telescopes.

\section*{Acknowledgments}

I thank David Jewitt (DJ) for insightful discussions and comments on the
manuscript, and for correctly predicting that I was going to write it.  I also
acknowledge Nuno Peixinho for helpful comments.  This work was supported by a
grant from NSF to DJ.


%
%
%
%
%
%
%
%
%

\end{document}